\documentclass[10pt,twocolumn,showpacs,preprintnumbers,amsmath,amssymb,prb]{revtex4-1}
\usepackage{graphicx}  
\usepackage{dcolumn}   
\usepackage{bm}        
\usepackage[utf8]{inputenc}
\usepackage{graphics}
\usepackage{latexsym}
\usepackage{amsmath}
\usepackage{float}
\usepackage{cancel}
\usepackage{multirow} 
\usepackage{pstricks,pst-plot}
\newcommand{\bra}[1]{\ensuremath{\langle #1 |}}
\newcommand{\ket}[1]{\ensuremath{| #1 \rangle}}
\newcommand{\bracket}[2]{\ensuremath{\langle #1 | #2 \rangle}}

\begin{document}


\title{Mass-profile quantum dots in graphene and artificial periodic structures}
\author{A. Guti\'errez-Rubio  and T. Stauber}
\affiliation{Departamento de Teor\'{\i}a y Simulaci\'on de Materiales, Instituto de Ciencia de Materiales de Madrid, CSIC, E-28049 Madrid, Spain}
\date{\today}

\begin{abstract}
We analyze the bound-state spectra of mass-profile quantum dots in graphene, a system at current experimental reach. Homogeneous perpendicular magnetic fields are also considered which result in breaking the valley degeneracy. The spectra show rich features, arising from the chiral band structure of graphene and its Landau levels and we identify three different regimes depending on the ratio between the radius of the dot and the magnetic length. We further carry out a comparison with potential-well quantum dots discussed in  [Recher {\it et al}, Phys. Rev. B {\bf 79}, 085407 (2009)] and conclude that mass confinement may offer significant advantages for optical applications in the THz and infrared regime. Also due to experimental advances, we additionally analyze the band structure of a linear chain of mass-profile quantum dots, where overlap-assisted hopping processes play a major role for closely packed arrays. The inclusion of Coulomb interactions between electron-hole pairs of adjacent sites leads to a new regime where F\"orster transfer processes become dominant.  
\end{abstract}
\maketitle


\section{Introduction}

Due to the chiral nature of its carriers, gapless graphene cannot confine electrons via a lateral electrostatic potential.\cite{Katsnelson06,neto2009electronic}
Several setups have been found to tackle this problem, producing nanostructures of graphene in form of quantum dots\cite{Ponomarenko08,Volk13} or nanoribbons.\cite{Stampfer09,baringhaus2014exceptional}
Also confining Dirac electrons in rings with edge reconstruction,\cite{recher2007aharonov,wunsch2008electron,romanovsky2013topological} inhomogeneous constant magnetic fields,\cite{silvestrov2007quantum,de2007magnetic,peres2009dirac} superlattices over different substrates\cite{rusponi2010highly,hunt2013massive,fuhrer2013critical} with a modulated Fermi velocity\cite{lima2014controlling} or scalar potential\cite{maksimova2012graphene} as well as nanohole patterning\cite{liu2009band} and topological mass terms\cite{jackiw2012fractional,ferreira2013magnetically,hasan2010colloquium,qi2011topological}  have been discussed. Achieving this gap opening is essential in regard to the design of nanodevices or possible applications to quantum computing.\cite{loss1998quantum,trauzettel2007spin}  

Once we count on the possibility of opening a gap and confining electronic states, the ability to control the level degeneracy is of high interest. As an example, valleytronic devices\cite{rycerz2007valley} or spin qubits\cite{trauzettel2007spin,Guttinger10} usually require a lifted valley degeneracy to be engineered, although some alternatives have been proposed recently.\cite{rohling2014hybrid} This, in turn, is achievable by means of magnetic fields\cite{recher2009bound,recher2007aharonov} or etching graphene ribbons with armchair boundaries.\cite{trauzettel2007spin}


In this paper, we study yet another alternative to confine electrons in graphene consisting in a position dependent gap. This possibility was first discussed in the context of infinite-mass boundaries by Berry and Mondragon and was mainly motivated by theoretical considerations.\cite{Berry87}
Other works have also analyzed the spectra of circularly shaped finite-mass profiles in the presence of electromagnetic fields.\cite{giavaras2010graphene,giavaras2011dirac,zhu2012confined}
Here, our renewed interest is based on recent experiments on graphene on top of a Ir(111)-substrate covered by Iridium clusters. Covered and uncovered regions show different particle gaps, but the substrate and the clusters leave graphene's linear spectrum almost unaffected.\cite{rusponi2010highly} Quantum dots confined by a finite mass-boundary can thus be designed with nanosize accuracy where the local change in the one-particle gap is introduced by the removal of substrate clusters selected at will by a STM-tip.\cite{martinez2014towards} So different quantum dot sizes and geometries are within reach and shall be discussed here.  

In particular, we will study the bound-state spectrum of a circular dot as a function of the radius and the magnetic field dependence. We will further establish a comparison between our system, i.e., mass-profile quantum dots (MP-QDs) and a potential-well quantum dot (PW-QD) previously characterized in Refs. \onlinecite{recher2009bound,Recher10}. We discuss their similar dependence on the dot size and the magnetic field, and also that the valley degeneracy splits proportionally to $B$ in both cases. On the other hand, we remark several differences between them, e.g., the spectrum of MP-QDs being particle-hole symmetric and less dense, thus being more susceptible to optical experiments. Dealing with a simpler level structure may be advantageous  also for applications, which endorses the interest in featuring MP-QDs. Finally, we relate the properties of the spectra obtained at high $B$-fields to the non-trivial Berry phase of $\pi$ in graphene. This gives rise to striking differences with respect to quantum dots hosted by other systems like a conventional 2D electron gas. We emphasize the novelty of this regime, which has not been addressed in the aforementioned related works and might be at experimental reach also in systems with high local strains.\cite{levy2010strain}

At last, again owing to the experimental feasibility to create periodic arrays of identical nano-structures and in regard to possible applications like those introduced in Ref. \onlinecite{wang2011nanopatterned}, a linear chain of MP-QDs is analyzed. Its band structure is calculated for a fixed radius $R$ as a function of the lattice parameter $D$, and the relevance of overlap-assisted hopping processes is discussed for closely packed arrays. Single-particle processes such as hopping or spontaneous decay rates are compared to excitonic processes.
Varying the distance between the dots allows the tuning to a Frenkel excitonic regime which can be described by a bosonic tight-binding Hamiltonian. 

The paper is organized as follows. In Sec. \ref{themodel}, we present the model for mass-profile quantum dots and its solution. In Sec. \ref{B0case}, we discuss the spectrum at $B=0$ for MP-QDs and PW-QDs, whereas in Sec. \ref{Bneq0case} we extend this analysis to $B\neq 0$. Sec. \ref{latticesofMP-QDsA} focuses on a MP-QD one-dimensional linear chain, and in Sec. \ref{latticesofMP-QDsB} the Coulomb interaction between electrons is included. We summarize our conclusions in Sec. \ref{conclusions}. In three appendices, we present details of the solution of the eigenvalue problem and define the tight-binding model for arbitrary arrays of MP-QDs.

\section{The model}
\label{themodel}

We shall consider graphene with a position dependent gap $2\Delta(r)$.
The Hamiltonian can be written as
\begin{align}
H_{\tau}=H_0+\tau\Delta(r)\sigma_z, \label{HtauB}
\end{align}
with
\begin{align}
&
H_0=v_F(\vec{p}+e\vec{A})\cdot\vec{\sigma}\;, \label{H0}\\
&
\vec{B}=\nabla\times\vec{A}=(0,0,B)\;.
\end{align}
$\vec{\sigma}=(\sigma_x,\sigma_y)$ and $\sigma_z$ are the Pauli matrices, $v_F=10^6\text{m/s}$ is the Fermi velocity, $e>0$ and the index $\tau=\pm 1$ labels the valleys.

The mass profile is considered to be circularly symmetric and steplike, $\Delta(r)=m_1v_F^2\Theta(r-R)+m_2v_F^2\Theta(R-r)$, where $R$ is the radius of the quantum dot.
It can be solved by exploiting the rotational invariance, $[J_z,H_\tau]=0$, with $J_z=-i\partial_\phi+\sigma_z/2$ the total angular momentum ---orbital plus lattice--- projected onto the $\hat{z}$ direction. So in polar coordinates, the eigenvectors can be written as
\begin{align}
\psi^\tau(r,\phi)=e^{i(j-1/2)\phi}
\left(
\begin{array}{c}
\chi^\tau_A(r)\\
\chi^\tau_B(r)e^{i\phi}
\end{array}
\right)\;, \label{wfB}
\end{align}
where $j$ is a half odd integer. With Eq. (\ref{wfB}), the eigenvalue problem of the Hamiltonian, Eq. (\ref{HtauB}), can be written as
\begin{align}
r^2\partial_r^2\chi_{\sigma}^{\tau}(r)+r\partial_r\chi_{\sigma}^{\tau}(r)=
(b^2r^4+a_{\sigma}r^2+n_{\sigma}^2)\chi_{\sigma}^{\tau}(r)\;, \label{eqsigma}
\end{align}
where $\sigma=\pm1$ corresponds to the A/B sublattice. We have also defined
\begin{align}
&
a_\sigma=2b(j+\sigma/2)-(E^2-\Delta^2)/(\hbar v_F)^2\;, \label{parameters1}\\
&
\Delta=mv_F^2\;,\;
n_\sigma=|j-\sigma/2|\;,\;
b=\frac{eB}{2\hbar}\;. \label{parameters2}
\end{align}
The solutions of Eq. (\ref{eqsigma}) are given in Apps. \ref{appenBneq0} and \ref{appenB=0} for non-zero magnetic field $B\neq 0$ and $B=0$, respectively. The matching conditions imposed on Eq. (\ref{wfB}) at $r=R$, also detailed there, yield the bound-states energies.
The eigenstates will be labeled as $\ket{\tau,j,n}$, $n$ ordering them in ascending absolute value of the energy.

A similar model was recently studied in Ref. \onlinecite{recher2009bound},
\begin{align}
H_{\tau}=H_0+\tau\Delta_0\sigma_z+U(r), \label{HRecher}
\end{align}
with a steplike electrostatic, rotationally invariant potential, $U=\Theta(R-r)U_0$ ($U_0<0$). Unlike in the case of a MP-QD, Eq. (\ref{HtauB}), the gap of a PW-QD, $\Delta_0$, is not position dependent, whereas $H_0$ is again given by Eq. (\ref{H0}).
The differences are sketched in the insets of Fig. \ref{SpectraB0}, which outline the spectrum of the two different quantum dots, considered throughout this work. 

The relation between the eigenfunctions of Eqs. (\ref{HtauB}) and (\ref{HRecher}) is discussed in Apps. \ref{appenBneq0} and \ref{appenB=0}, as well as the symmetries they display.
Further discussion about the spectra is held in the subsequent sections.


The mass profile in graphene can be achieved when placing it on a Ir(111)-substrate and subsequently covering graphene with Ir or W-clusters. Graphene on Ir(111) displays a small gap of $\sim50$meV due to the Moir\'e-lattice structure formed by the graphene layer with the substrate, defining a super-lattice constant of 2.5nm. Owing to a change in the graphene lattice structure from sp${}^2$ to sp${}^3$-bonding in the covered region,\cite{Feibelman08} this gap is augmented  from 50meV to 400meV. In Ref. \onlinecite{martinez2014towards}, it was demonstrated that the upper metal clusters of the size of the Moir\'e-superlattice can be removed by an STM-tip at will, thus opening up the possibility of creating mass-confined quantum dots of arbitrary size up to nanometer accuracy. 
Throughout the article, we will therefore use the values found in photoemission spectroscopy experiments,\cite{rusponi2010highly} i.e.,
$\Delta_1=m_1v_F^2=0.025\;\text{eV}$ and $\Delta_2=m_2v_F^2=0.2\;\text{eV}$.
In the PW-QD, we choose a confining potential $U_0=(\Delta_1-\Delta_2)\Theta(r-R)$ which guarantees the the same well depth as in the MP-QD.

\section{Electronic spectrum of a single quantum dot}

\subsection{$B=0$ case}
\label{B0case}

\begin{figure}
\begin{center}
\includegraphics{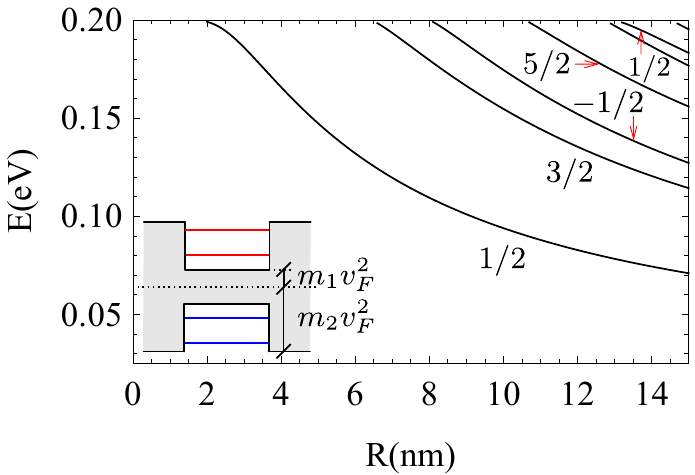}
\includegraphics{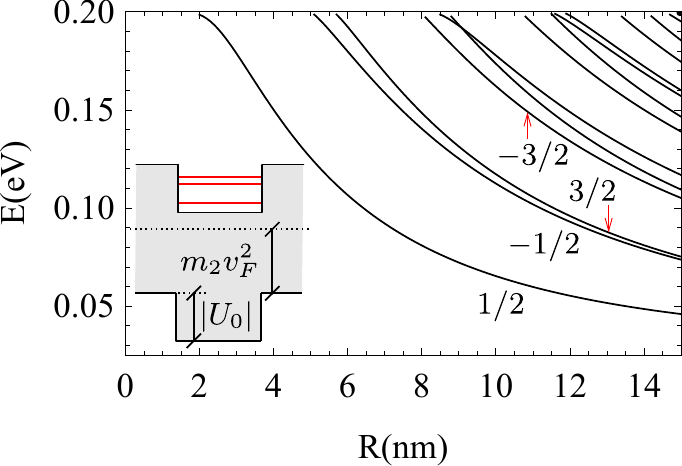}
\end{center}
\caption{(color online). Spectra of a circular MP-QD (upper panel) and a PW-QD (lower panel) for $B=0$. The values of $j$ respective to $\tau=1$ label the most bounded energy levels.
Insets: diagrams of the bound-state levels in the mass and potential well. The shadowed region corresponds to forbidden values of the energy, and the colored lines inside the wells represent bound-states. The electron (red)-hole (blue) symmetry is only present in the MP-QD.}
\label{SpectraB0}
\end{figure}

Let us first discuss the spectra in the absence of magnetic fields.
Results are plotted in Fig. \ref{SpectraB0}.
As it is proved in App. \ref{appenB=0}, the levels are doubly degenerate,
\begin{align}
E(\tau,j)=E(-\tau,-j), \label{symvalley}
\end{align}
reflecting the time reversal symmetry that connects the two valleys.
Only for MP-QDs the electron-hole symmetry is also present,
\begin{align}
E(\tau,j)=-E(-\tau,j), \label{symeh}
\end{align}
while absent in the case of PW-QDs (cf. also the discussion in App. \ref{appenB=0} and the insets of Fig. \ref{SpectraB0}).
This aspect implies a striking difference: MP-QDs can host confined electron-hole pairs, whereas they are not present in PW-QDs.

Further comparing the spectra of MP-QDs and PW-QDs, one can notice a higher confinement of the states of the latter. This implies a denser spectrum of PW-QDs for the same depth of the mass and potential well. In turn, the larger level spacing makes MP-QDs more accessible for optical spectroscopy.

For experimentally realizable MP-QDs with $R\lesssim 10\,\text{nm}$, intra-band transitions between bound-state levels have a frequency of the order of $10\,\text{THz}$, enabling our system to support terahertz optical applications. For MP-QDs with $R\lesssim 6\,\text{nm}$, only one bound-state in the conduction band is present, defining a possible qubit which can be optically turned on (electron-hole pair creation) and off (neutral ground state).  Furthermore, excitonic effects lifting the valley degeneracy could manifest many body effects also in the $\text{THz}$ range, promoting the system as an experimentally realizable probe for interactions in quantum dots, including electron-phonon coupling.\cite{Stauber06}

\subsection{$B\neq 0$ case}
\label{Bneq0case}

\begin{figure}
\begin{center}
\includegraphics{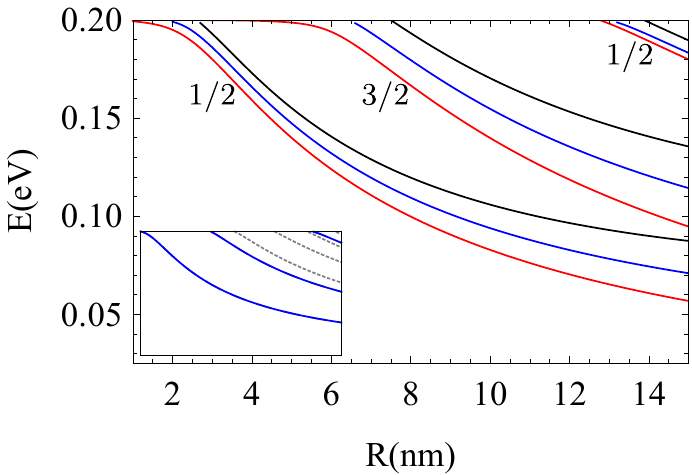}
\includegraphics{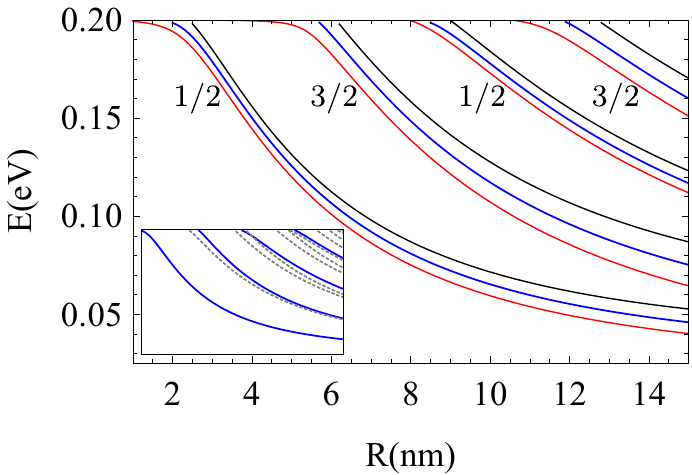}
\end{center}
\caption{(color online). Splitting of the valley degeneracy of a circular MP-QD (upper panel) and a PW-QD (lower panel) when a perpendicular magnetic field is applied. Blue lines correspond to $B=0$, and they split into black ($\tau=1$) and red ($\tau=-1$) lines when applying $B=4\text{T}$. For clarity, only levels with $|j|=\{1/2,3/2\}$ (labels of the curves) have been plotted. In the insets, where the full spectra at $B=0$ is shown, they are highlighted in blue.
}
\label{Splitting}
\end{figure}

In Fig. \ref{Splitting}, we show the spectra of a MP-QD (upper panel) and a PW-QD (lower panel) as a function of the radius for zero and non-zero magnetic field $B$. We observe a splitting between levels belonging to different ($\tau=\pm 1$) valleys which is proportional to the magnitude of the applied magnetic field.
For $B=4\,\text{T}$, we can achieve a splitting up to $2\,\text{meV}$ (THz regime) for quantum dots with only one bound-state (i.e., for $R\lesssim 6\,\text{nm}$).
Remarkably, the splitting of the levels is considerably larger in the MP-QD.


These plots also show that if levels belong to different valleys, their energies are modified by the magnetic field in a different manner. Whereas every $\tau=+1$ level rises with $B$, some $\tau=-1$ ones are lowered, i.e., become more confined.
Interestingly, for some $R$ values for which there were an equal number of $\tau=\pm 1$ states at $B=0$, we observe that new $\tau=-1$ states appear and other $\tau=+1$ states vanish when applying a sufficiently high magnetic field. The opposite happens in the valence band. As we will see later, this fact will be relevant to explain the spectrum for $R\gg l_B$, with the magnetic length $l_B=\sqrt{\hbar/(eB)}$.

\begin{figure*}
\begin{center}

\includegraphics{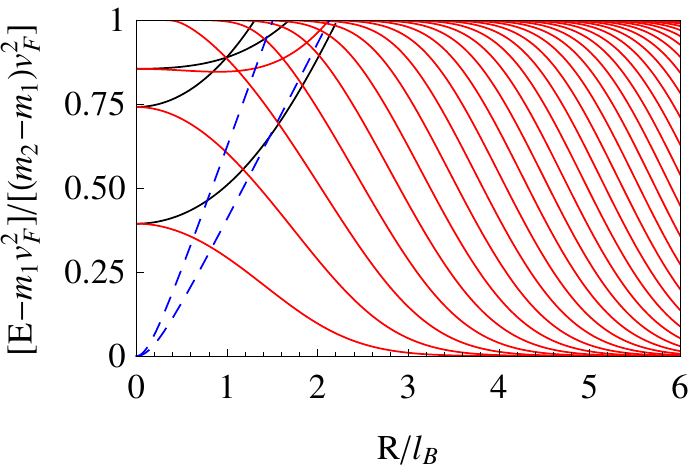}\hspace{1cm}
\includegraphics{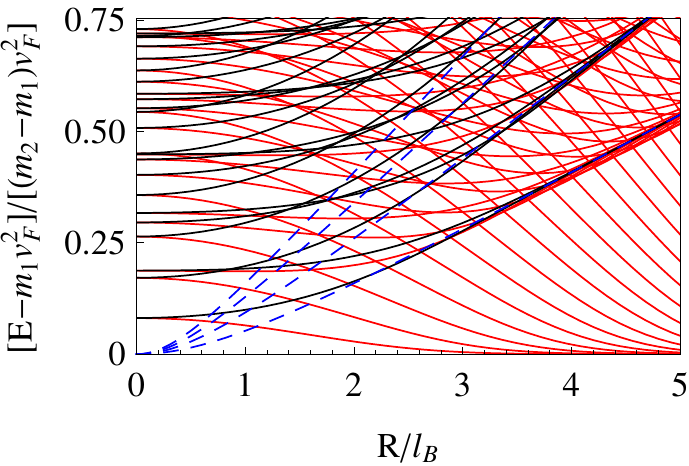}
\includegraphics{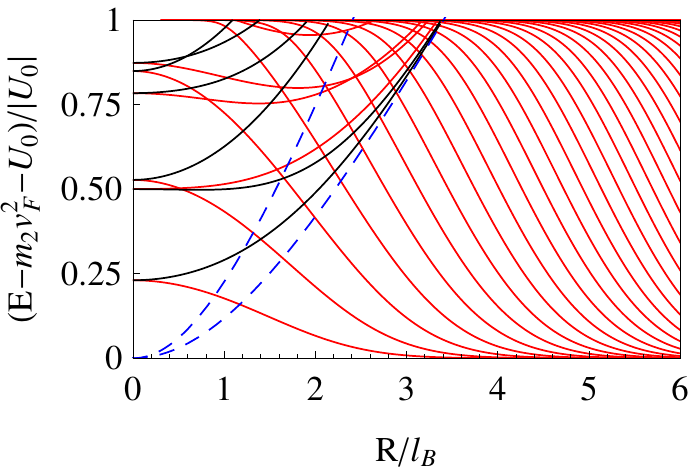}\hspace{1cm}
\includegraphics{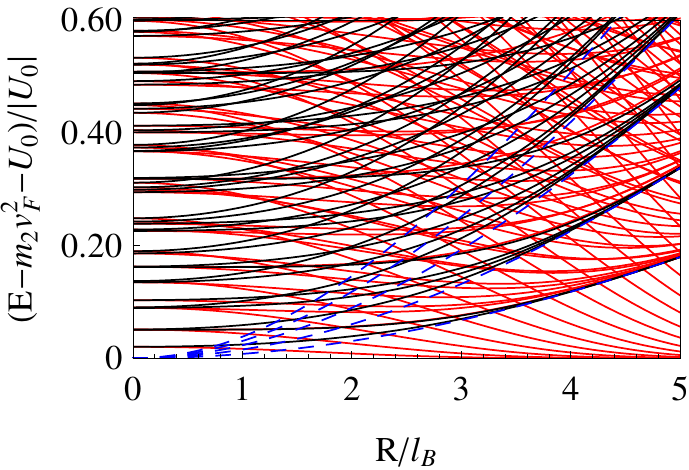}
\end{center}
\caption{(color online). Upper panels: MP-QDs. Lower panels: PW-QDs. Left panels: spectra for $R=10\;\text{nm}$ ($R=l_B\Rightarrow B=6.6\,\text{T}$). Right panels: spectra for $R=40\;\text{nm}$ ($R=l_B\Rightarrow B=0.41\,\text{T}$). Black (red) corresponds to $\tau=(-)1$ valley. Convergence to Landau levels (bottom of the well and dashed blue lines) can clearly be appreciated.
The spectrum structure with only $\tau=-1$ levels and approximately equidistant energy levels appears for $B>B_c$, see Eq. (\ref{criticalB}).}
\label{SpectrumB}
\end{figure*}



In Fig. \ref{SpectrumB}, we present the spectra as a function of the fixed dot radius $R$ over the magnetic length $l_B$. They show more clearly one of our previous considerations concerning the appearance and disappearance of states with different valley index, see Fig. \ref{Splitting}, i.e., as long as $B$ is increased for a fixed $R$, some $\tau=+1$ levels are no longer bound-states whereas others with $\tau=-1$ enter the well. Remarkably, one can appreciate three (and not two, as usual) different regions in these plots for both the MP-QD and the PW-QD: i) for $B\to 0$, we recover the degenerate spectrum, in which the level spacing does not seem to have a definite structure. ii) As soon as $B$ is sufficiently large, the levels converge to straight lines and to the bottom of the well. iii) For greater values of $B$, only bound-states with $\tau=-1$ are allowed in the conduction band, see left panels of Fig. \ref{SpectrumB}. In this last case, for energies not close to the top or the bottom of the well, we encounter equally spaced non-degenerate levels.
Near the bottom, we can notice how a growing number of states with $B$ converge to the lowest possible energy.
We will devote part of this section to discuss the emergence of these patterns featured in points ii) and iii).

Concerning the intermediate region ii), Fig. \ref{SpectrumB} resembles a typical Fock-Darwin spectrum,\cite{giuliani2005quantum,ferreira2013magnetically} the potential well or the mass profile playing the role of the harmonic potential in that model. In this regime, the energy term corresponding to $B$ dominates over the well depth. This is confirmed by the dependence
\begin{align}
|E_n|=\sqrt{\Delta^2+2n\left(\hbar v_F/l_B\right)^2}
\end{align}
($n\geq 0$ refers to the $n$th asymptote), revealing a structure characteristic of Landau levels in gapped graphene.\cite{haldane1988model,koshino2010anomalous} The existence of the lowest Landau level is of particular relevance, since it is responsible for the structure of the spectrum for large $B$, as discussed below.

The degeneracy of the Landau levels in this region is also noteworthy, which is proportional to the magnetic flux through the system. One can verify this dependence in Fig. \ref{SpectrumB}: while $B$ is increased, more $\tau=-1$ levels appear inside the wells and converge asymptotically to graphene Landau levels, providing them with the required degeneracy.

For $B$ greater than a critical value
\begin{align}
B_c\simeq \frac{\left(m_2^2-m_1^2\right)v_F^2}{2e\hbar}\simeq 30\; \text{T}, \label{criticalB}
\end{align}
only the lowest Landau level will remain, see Fig. \ref{SpectrumB}.
We encounter the aforementioned region iii), with only $\tau=-1$ states. Our previous considerations on the degeneracy explain the level structure of this part of the spectrum. As long as $B$ is increased, a constant income of levels is needed to guarantee that the lowest Landau level is degenerate enough. As a result, our quantum dots show an excited spectrum of equally spaced levels that will converge to the bottom of the well at higher $B$. Their difference in energy can be tuned with the radius, since a higher area $\mathcal{A}$ increases the degeneracy of Landau levels, in turn implying a greater density of incoming states.
Experiments with graphene have been carried out for values of $B\gtrsim B_c$,\cite{plochocka2008high} so this regime may be observable for our values of $\Delta_1$ and $\Delta_2$.
As an alternative, pseudo-magnetic fields exceeding $B_c$ could be induced by strain.\cite{levy2010strain}

As we have seen, the chiral nature of graphene's carriers manifests itself in the results. The presence of a Landau level whose energy is $B$ independent is the most determinant feature in the quantum dot spectrum. It guarantees the existence of bound-states at arbitrarily high values of $B$ with the structure of the third region discussed before.
This is in stark contrast to quantum dots of ordinary 2D semiconductors, where no bound-states exist beyond some critical $B_c$. The approximate equidistant level structure might be useful for optical experiments in the THz-regime, inducing transitions between several adjacent levels. Since conduction and valence bands host bound-states of opposite valleys, no interband transitions are allowed.

\section{Arrays of MP-QDs}

\label{latticesofMP-QDs}
The controlled removal of metal clusters on top of graphene placed on a Ir(111)-substrate allows for creating artificial periodic lattice structures within the nanoscale.\cite{martinez2014towards} In this section, we will thus focus on linear chains of MP-QDs, setting with this elementary example the proceeding to analyze more complicated one- or two- dimensional arrays. For the sake of simplicity, we fix the dot radius $R=6.5\,\text{nm}$, which implies dealing only with a single bound-state (cf. Fig. \ref{SpectraB0}) per band. Our aim is to carry out a tight binding calculation considering both the valence and the conduction band. Later on, we will also add the Coulomb interaction between the excitations hosted in different dots. 

\subsection{One-particle physics}
\label{latticesofMP-QDsA}
We start our analysis by considering two MP-QDs whose centers are separated by a distance $D$. It turns out that the overlap $\lambda$ of the two wavefunctions is not negligible for $D$ close to $2R$. This is demonstrated in Fig. \ref{Solape}. Although its square is smaller, we will not neglect it for the moment. Actually, we will show below that its contribution will be relevant for an array of quantum dots with a small lattice parameter $D\simeq 2R$. Henceforth, only $O(\lambda^n)$-terms with $n\geq3$ will be discarded unless the contrary is specified. This introduces an error of less than 1\%, see lower panel of Fig. \ref{Eorders}.

\begin{figure}
\begin{center}
\includegraphics{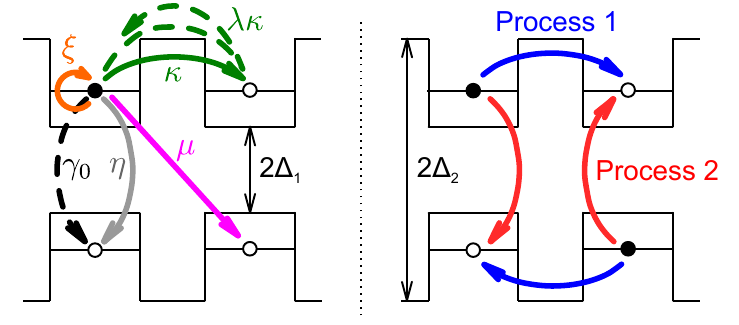}
\end{center}
\caption{(color online). Outline of different hopping processes between two MP-QDs. Left hand side: single-particle processes. $\gamma_0$ is the spontaneous decay, Eq. (\ref{gamma0}). Right hand side: interaction processes.}
\label{pozoshoppings}
\end{figure}

\begin{figure}
\begin{center}
\includegraphics{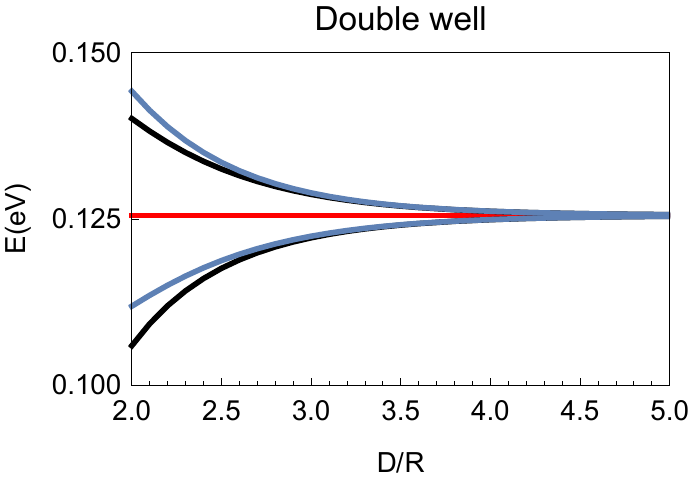}
\includegraphics{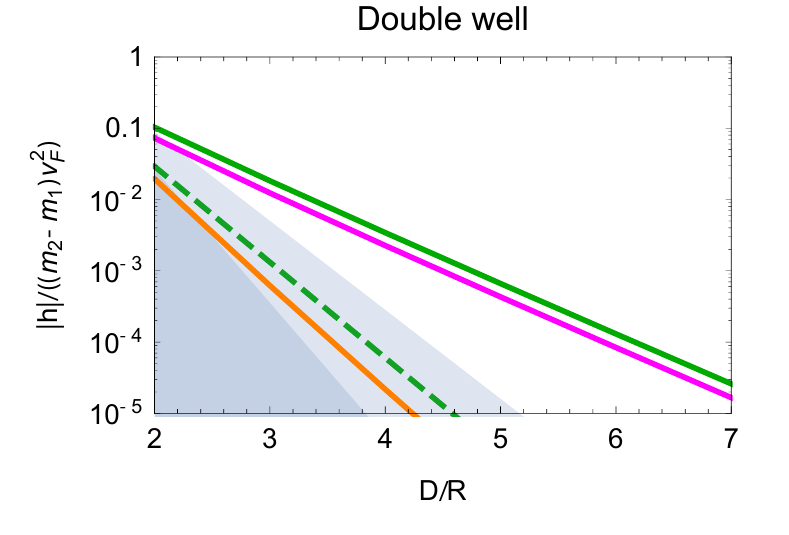}
\end{center}
\caption{(color online). Upper panel: Bound-state energies of two nearby MP-QDs. Red: Bound-state energy of a single well. Black (blue): Bound-state energies neglecting the overlap squared (cubed) of the wavefunctions. For more details, see App. \ref{appenB}.
Lower panel: Comparison between the hopping processes depicted in Fig. \ref{pozoshoppings} (with the same color code). $h$ labeling the vertical axis refers to $\kappa$, $\mu$ and $\xi$. The strong (light) shadowed region corresponds to values below the overlap cubed (squared).}
\label{Eorders}
\end{figure}

A sketch of all different processes that can take place between the two MP-QDs is shown in Fig. \ref{pozoshoppings}. In App. \ref{appenB}, we demonstrate that the probability amplitudes of the hopping-processes $\xi$, $\kappa$, $\mu$ and $\lambda\kappa$ are relevant, whereas $\eta$, the electron-hole annihilation, can be shown to be precisely zero. 
$\lambda\kappa$ is a term of the order of $\lambda^2$ (cf. Fig. \ref{Eorders}).
It is not a direct hopping parameter like $\{\eta,\kappa,\mu\}$, but a hopping process provoked or \emph{assisted} by the overlap.

The upper panel of Fig. \ref{Eorders} shows the spectrum of the double well as a function of the distance between the centers.
There, it can be seen that considering or neglecting terms of the order of the overlap squared does not alter the values of the energies considerably.
Nevertheless, the situation will be different in periodic arrays of quantum dots closely packed, as we will discuss below.

In App. \ref{appenB}, we define a tight binding model for a linear chain of MP-QDs with $R=6.5\,\text{nm}$.
The resulting bands appear in Fig. \ref{Figbandas}, ranging from dispersive to flat with the increase of $D$. The gaps and carrier effective masses are thus tunable with the lattice parameter.

Clearly, for a lattice parameter close to the dot diameter, the effect of the overlap on the band structure is significant.
This is mainly due to the next-nearest neighbor hopping assisted by $\lambda$, see Fig. \ref{neighbors} and its discussion in App. \ref{appenB}.
Note that this effect cannot take place in a double well and this explains why the influence of the overlap-assisted processes on the eigenenergies was much weaker.

As a result, the effective mass $m^*$ of the carriers with $k\simeq 0$ is strongly renormalized by $\lambda$.
Remarkably, for a closely packed chain, the overlap-assisted processes give rise to a change of sign in the curvature of the bands around $k=0$ (see inset of Fig. \ref{Figbandas}).
For low densities, the ground state is thus given by a Fermi ring\cite{stauber2007fermi} and shows that the implications of considering second order processes in a tight binding approach go beyond a mere correction in eigenenergies.

\subsection{Coulomb interaction}
\label{latticesofMP-QDsB}
The second part of this section aims to include Coulomb interactions in our system. Processes like those depicted on the right hand side of Fig. \ref{pozoshoppings} come into play.
Their rates $\gamma_C$ can be calculated with the help of Fermi's golden rule,
\begin{align}
\gamma_C=\frac{2\pi}{\hbar}|\langle f|V_{\text{int}}|i \rangle|^2 F_{fi}\;, \label{Eq:t}
\end{align}
where $V_{\text{int}}$ is the Coulomb interaction and $F_{fi}$ the generalized delta-function, see Eq. (\ref{overlap}).\cite{Govorov03} $\ket{i}$ ($\ket{f}$) refers to the initial (final) state of the transition which is a two-particle ---electron-hole--- state
\begin{align}
|E\vec{R}_a\rangle \otimes |E'\vec{R}_b\rangle\;, \label{doubleparticle}
\end{align}
where, in turn, $|E\vec{R}_\alpha\rangle$ describes the eigenstate with energy $E$ of a single well centered at $\vec{R}_\alpha$. Since we are dealing with only a single bound-state per well in the valence and conduction band, and since the Coulomb interaction does not couple valleys nor spins, we can neglect the other quantum numbers ($j$, $\tau$, spin) that strictly label the state and only keep $E$ and $\vec{R}_\alpha$ in Eq. (\ref{doubleparticle}). In the overlap factor
\begin{align}
\label{overlap}
F_{fi}=\int_{-\infty}^{\infty} d\epsilon\,\rho_i(\epsilon)\rho_f(\epsilon)\;,
\end{align}
a Lorentzian was considered for the density of states of the $\ket{i}$ ($\ket{f}$) level, whose energy is centered at $E_{i(f)}$:
\begin{align}
\rho_{i(f)}(\epsilon)=\frac{1}{\pi}\frac{\Gamma}{(\epsilon-E_{i(f)})^2+\Gamma^2}\;.
\end{align}
A typical broadening of $\Gamma=10\,\text{meV}$ was used.\cite{mak2008measurement}

\begin{figure}
\begin{center}
\includegraphics{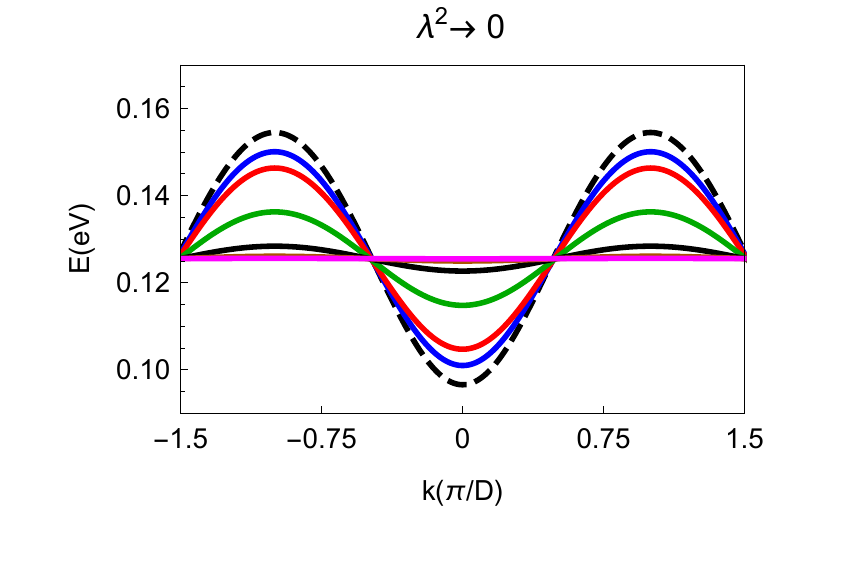}
\includegraphics{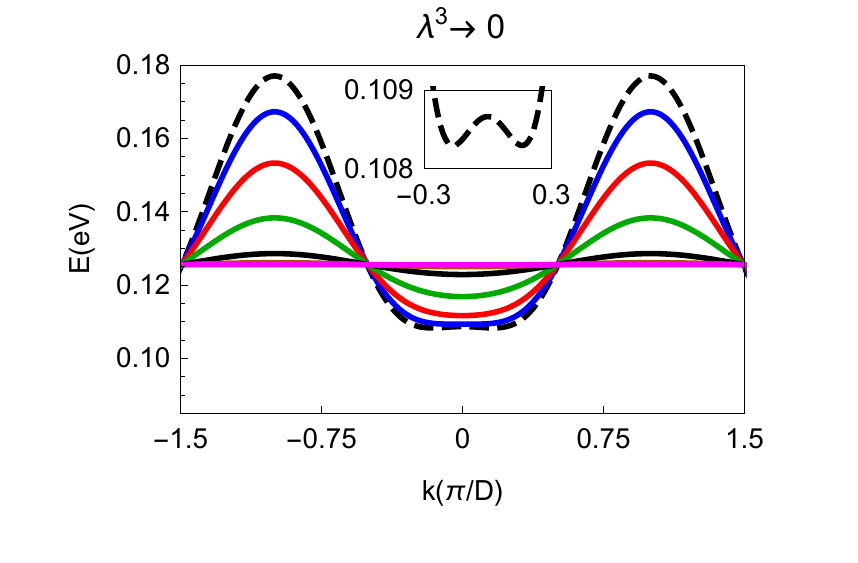}
\end{center}
\caption{(color online). Bands of a one-dimensional chain of quantum dots. Dashed black: $D=2\,R$; blue: $D=2.1\,R$; red: $D=2.3\,R$; green: $D=2.7\,R$; solid black: $D=3.5\,R$; orange: $D=4.5\,R$; magenta: $D=6\,R$. Terms of higher order than the overlap squared (cubed) are neglected in the upper (lower) panel.
The inset plots a close view of the $D=2\,R$ curve, showing its negative curvature at $k=0$.}
\label{Figbandas}
\end{figure}

Process 2 in Fig. \ref{pozoshoppings} is an example of F\"orster transfer.\cite{novotny2012principles}
In this case, one can approximate the matrix element in Eq. (\ref{Eq:t}) by its multipolar expansion,
\begin{align}
\langle f |V_{\text{int}}| i \rangle \simeq
\frac{1}{4\pi\epsilon_0}
\frac{D\vec{\mu}_a\cdot\vec{\mu}_b-3(\vec{\mu}_a\cdot\vec{D})(\vec{\mu}_b\cdot\vec{D})}{D^5}\;, \label{Eq:tF}
\end{align}
where $\vec{D}=\vec{R}_a-\vec{R}_b$, and in turn $\vec{R}_a$ and $\vec{R}_b$ are the centers of the MP-QDs involved in the process.
The dipole momenta are
\begin{align}
\vec{\mu}_j=\int d^2r\, |\psi(\vec{r}-\vec{R}_j)|^2(\vec{r}-\vec{R}_j)\;,\;\ j=\{a,b\}\;, \label{mu}
\end{align}
with $|\psi(\vec{r}-\vec{R}_j)|^2$ the density probability associated to the single-particle bound-state centered at $\vec{R}_j$.
We checked numerically the excellent agreement between Eq. (\ref{Eq:tF}) and the exact value of the transition matrix element.

\begin{figure}
\begin{center}
\includegraphics{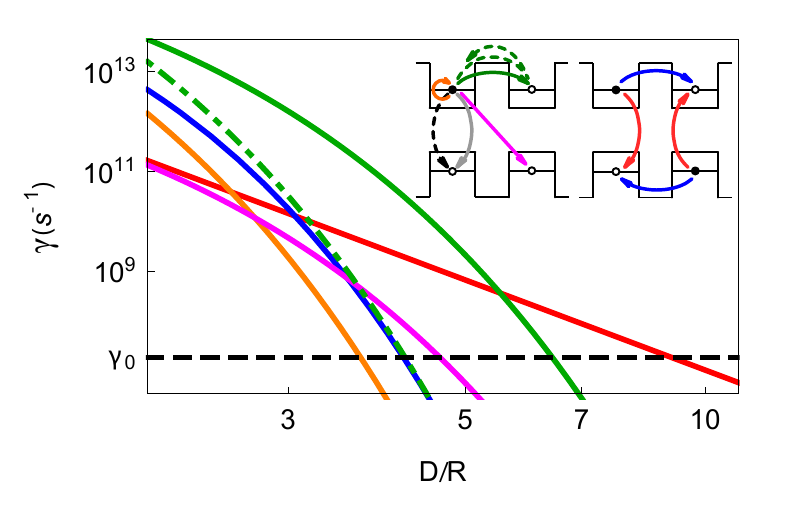}
\end{center}
\caption{(color online). Comparison between the rates $\gamma$, Eq. (\ref{Eq:t}), of single-particle and excitonic processes. The color code was introduced in Fig. \ref{pozoshoppings}. It is reproduced in the inset for clarity.}
\label{processes}
\end{figure}

A comparison between the rates $\gamma$ of all processes depicted on Fig. \ref{pozoshoppings} is shown in Fig. \ref{processes}.
We highlight the algebraic behavior of the F\"orster transfer versus the exponential one of all the rest.
The spontaneous decay rate,\cite{novotny2012principles}
\begin{align}
\gamma_0=\frac{\omega_0^3|\vec{\mu}|^2}{3\pi \epsilon_0\hbar c^3}\simeq 1.9\cdot 10^7\;\text{s}^{-1}\;, \label{gamma0}
\end{align}
is also plotted as reference, with $\vec{\mu}$ given by Eq. (\ref{mu}) and $\hbar\omega_0$ being the energy difference between the levels involved.
For $D\simeq 7R$, this indicates the existence of a regime in which F\"orster transfer is the dominant process. In that case, the particle-hole excitations have a sufficiently long lifetime to overlap with the adjacent site to form a band. These Frenkel excitons can thus be described by the following quasi-bosonic tight-binding Hamiltonian in the diluted limit:
\begin{align}
&
H\simeq
-\sum_{\langle i,j \rangle}
\left[
t_{ex}
a^\dagger(\vec{R}_j)a(\vec{R}_i)+\mathrm{h.c.}
\right].\label{Hboson}
\end{align}
Here,  $a(\vec{R}_i)\equiv c^\dagger_{-}(\vec{R}_i)c_{+}(\vec{R}_i)$ annihilates and $a^{\dagger}(\vec{R}_i)\equiv c^\dagger_{+}(\vec{R}_i)c_{-}(\vec{R}_i)$ creates an exciton at lattice site $\vec{R}_i$, where $c^{(\dagger)}_+(\vec{R}_i)$ and $c^{(\dagger)}_-(\vec{R}_i)$ are the electron annihilation (creation) operators in the upper and lower levels of a single dot centered at $\vec{R}_i$, respectively. The excitonic operators $a^{(\dagger)}(\vec{R}_i)$ satisfy bosonic commutation relations in the diluted limit.\cite{solyom2010fundamentals} 

The effective excitonic hopping amplitude induced by F\"orster transfer is given by $t_{ex}\equiv\langle +\vec{R}_j;-\vec{R}_i|V_{int}|{-\vec{R}_j};+\vec{R}_i\rangle$, which takes place only between nearest neighbors. Other processes, outlined in App. \ref{appenB} and which were neglected in Eq. (\ref{Hboson}), would induce a finite lifetime of the excitons. For lattices with $D/R\lesssim 3.5$, the bands turn dispersive (see Fig. \ref{Figbandas}) and single-particle processes become dominant.

The previous considerations promote the system under study to a highly tunable probe which can further host collective excitations in form of interband plasmons.\cite{Stauber14,langer2011sheet} The engineering of lattices exhibiting different symmetries and dimensionality thus opens up a new scenario to explore interactions in artificial lattices.

\section{Conclusions}
\label{conclusions}

Motivated by recent experimental advances, we have studied the bound-state spectra of mass-profile quantum dots (MP-QDs) and compared it with the corresponding spectrum of recently studied potential-well quantum dots (PW-QDs). Both systems allow to confine electrons in 2D and control the lifting of the valley degeneracy by applying a perpendicular magnetic field to the sheet. We have seen that the behavior of their spectra as a function of the radius and the magnetic field is similar, but the level structure of MP-QDs is somehow simpler. This could make the latter more suitable for optical applications in the mid-infrared and $\text{THz}$ regime. Moreover, we have featured different regions of the spectra according to the magnitude of the magnetic field. Besides the quantum dot spectrum at $B\to 0$, we have discussed the convergence to Landau levels at intermediate values of $B$ and the appearance of an equally spaced level structure for large $B$-fields. The latter arises due to the existence of a lowest Landau level pinned to the band edge whose energy is $B$ independent, in stark contrast with quantum dots hosted by conventional 2D systems.

In the second part, the electronic spectrum of linear chains of MP-QDs with fixed radius $R=6.5\,\text{nm}$ (and therefore a single valence and conduction bound level) was discussed.
Bands of tunable gap and curvature were then obtained.
Overlap-assisted processes are shown to play a significant role for closely packed arrays, renormalizing the effective mass of the carriers to the extent of changing its sign for $D\simeq 2R$.
When including interactions, the engineered lattice parameter allows to encounter Frenkel excitons.
Remarkably, a system of bosons in a linear chain ultimately governed by efficient F\"orster transfer between adjacent dots can be reproduced for $D\simeq 7R$.
Further research on the possibility of hosting Bose Einstein condensates in this kind of systems remains to be explored in future works.


\section{Acknowledgements}

We thank H. Ochoa, B. Amorim, F. Marchetti, I. Brihuega and F. Guinea for useful discussions. This work was supported by Spain's MINECO under Grant No. FIS2013-44098-P. 

\appendix

\begin{widetext}
\section{Wavefunctions of the bound-states for $B\neq 0$}
\label{appenBneq0}

In this appendix, we present the general formulas necessary to solve the eigenvalue problem of a MP-QD. The general solution of Eq. (\ref{eqsigma}) is given by
\begin{align}
&
\chi_\sigma^{\tau}(r)
=
e^{-br^2/2}r^{n_{\sigma}}
\left\{
\begin{array}{cc}
\alpha_{\sigma}M\left[q_{\sigma}(m_1),1+n_{\sigma},br^2\right] & \text{for } r\leq R\;,\\
\beta_{\sigma}U\left[q_{\sigma}(m_2),1+n_{\sigma},br^2\right] & \text{for } r>R\;,
\end{array}
\right. \label{generalsolution}
\end{align}
with
\begin{align}
&
q_\sigma(m)=\frac{1}{4}\left[\frac{a_\sigma}{b}+2(1+n_\sigma)\right]\label{qsigma}
\end{align}
and Eqs. (\ref{parameters1}) and (\ref{parameters2}). $U$ and $M$ are the confluent hypergeometric functions.\cite{abramowitz1970handbook}
Since the wavefunction has to be non singular at the origin and square integrable, $U$ ($M$) can only be a solution in the region $r>R$ ($r\leq R$).
The ratio $\alpha_{\sigma}/\beta_{\sigma}$ is provided by the coupled equations resulting from inserting Eq. (\ref{wfB}) in Eq. (\ref{HtauB}).

When imposing the continuity of the wavefunction at the frontier $r=R$,\cite{alonso1997on,mccann2004symmetry} the following conditions are obtained:
\begin{align}
&
j>0\Rightarrow \frac{U(q_1(m_2),1+n_1,bR^2)}{M(q_1(m_1),1+n_1,bR^2)}
=
\frac{\tau E+\Delta(m_1)}{[\tau E+\Delta(m_2)]\left(1-\frac{q_1(m_1)}{1+n_1}\right)}
\frac{U(q_{-1}(m_2),1+n_{-1},bR^2)}{M(q_{-1}(m_1),1+n_{-1},bR^2)}\;,
\label{EboundBap}
\\
&
j<0\Rightarrow\frac{U(q_{-1}(m_2),1+n_{-1},bR^2)}{M(q_{-1}(m_1),1+n_{-1},bR^2)}
=
-\frac{(1+n_{-1})q_{-1}(m_2)[\tau E-\Delta(m_1)]}{q_{-1}(m_1)[\tau E-\Delta(m_2)]}
\frac{U(q_{1}(m_2),1+n_{1},bR^2)}{M(q_{1}(m_1),1+n_{1},bR^2)}\;.
\label{EboundBap2}
\end{align}
They yield the allowed energies of the bound-states.

It is possible to relate the solution of Eq. (\ref{eqsigma}) (MP-QDs) and Eq. (\ref{HRecher}) (PW-QDs).
The substitution
\begin{align}
E\to E-U(r)\;,\; \ \Delta\to\Delta_0\;, \  m_1\to m_2\;,
\end{align}
in Eq. (\ref{generalsolution}) provides the wavefunctions of Eq. (\ref{HRecher}).
The matching conditions then yield Eqs. (\ref{EboundBap}) and (\ref{EboundBap2}) but with the changes
\begin{align}
&
\tau E\pm\Delta(m_1)\to \tau (E-U_0)\pm\Delta_0\;,\label{change1}\\
&
\tau E\pm\Delta(m_2)\to \tau E\pm\Delta_0\;. \label{change2}
\end{align}
Moreover, in Eq. (\ref{parameters1}) when inserted in Eq. (\ref{qsigma}), $E\to E-U_0$ for $q_{\sigma}(m_1)$ whereas $E\to E$ for $q_{\sigma}(m_2)$.
The whole set of substitutions can be understood under the following consideration: focusing on Eqs. (\ref{EboundBap}) and (\ref{EboundBap2}), the different masses $m_1$ and $m_2$ only appear for the regions $r\leq R$ and $r>R$, respectively. In order to obtain the solution of a PW-QD, we thus change $E\to E-U_0$ only in the case $r\leq R$. On the other hand, $m_1\to m_2$, i.e., $\Delta(r)\to\Delta_0$, holds everywhere.

The electron-hole symmetry of the solution for MP-QDs can be inferred directly from Eqs. (\ref{EboundBap}) and (\ref{EboundBap2}). They depend on the energy solely through the variable $\tau E$.
Therefore, given a state with $(E,\tau,j)$, another with $(-E,-\tau,j)$ exists.
Nevertheless, this symmetry is broken in PW-QDs.
The reason is that the substitutions given by Eqs. (\ref{change1}) and (\ref{change2}) in Eqs. (\ref{EboundBap}) and (\ref{EboundBap2}) split the dependence on $\tau E$ in $\tau$ and $E$ separately.

Plots of the wavefunction components appear in Fig. \ref{wfwithB}. In the upper panels, the effect of increasing the magnetic field is analyzed. A polarization of the B-sublattice takes place in both valleys as well as the quenching of the kinetic energy with magnetic field is revealed as a shift of the radial probability towards the center of the well. The lower panels, on the other hand, focus on levels with different $j$ values at a fixed $B$-field. The increase in the total angular momentum $J_z$ entails a shift of the radial probability away from the center of the dot. This is in agreement with the lower localization of the states with energies closer to the top of the well.

\begin{figure}
\begin{center}
\includegraphics{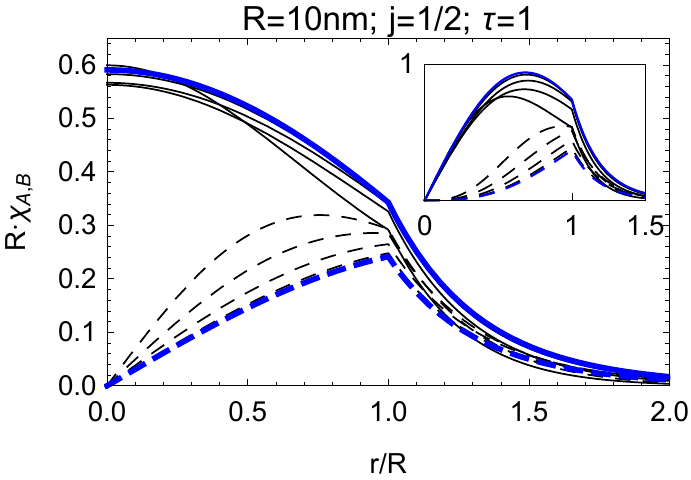}
\hspace{1cm}
\includegraphics{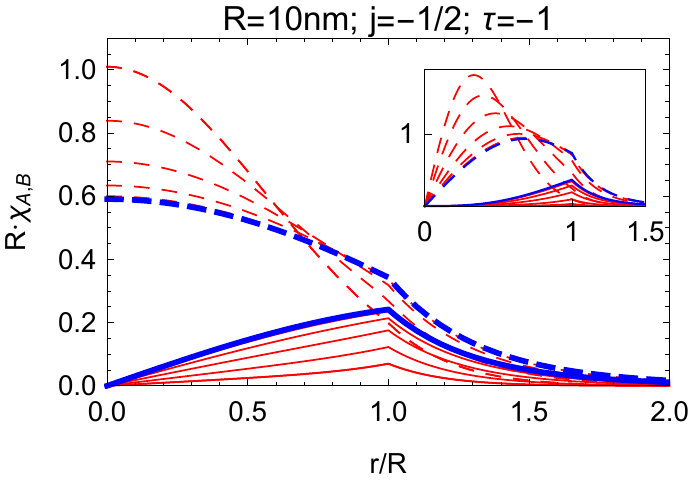}
\includegraphics{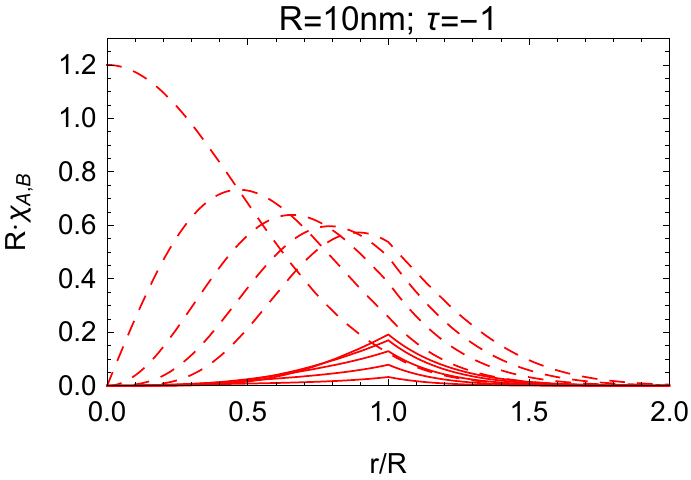}
\hspace{1cm}
\includegraphics{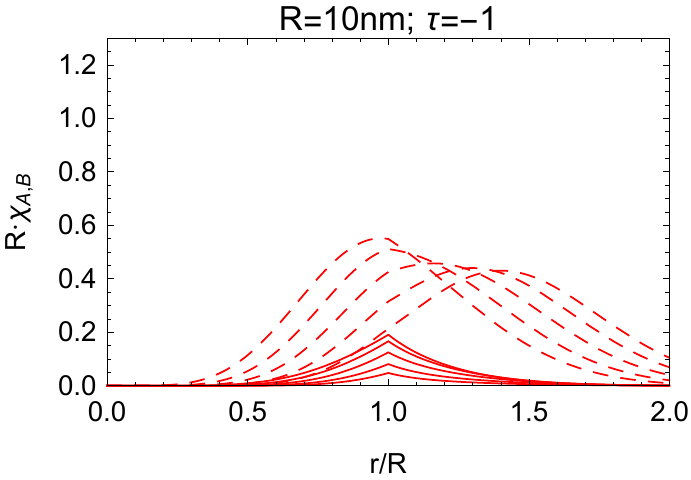}
\end{center}
\caption{(color online). Components of the MP-QD wavefunction, Eqs. (\ref{generalsolution}) and (\ref{wfB}). Solid (dashed) lines refer to the A (B) sublattice.
Upper panels: Plots for different values of the magnetic field $B$. The blue curves correspond to zero magnetic field ($B=0$), whereas black ($\tau=1$) and red ($\tau=-1$) curves show the eigenfunctions at finite magnetic field for $R/l_B=\{1,1.5,2\}$: greater values of $R/l_B$ are respective to the more deviated curves from the blue ones. For comparison, an extra solution for $R/l_B=2.5$ in the regime with only $\tau=-1$ levels has been plotted for $\tau=-1$. The insets depict the radial probability ($2\pi rR|\chi_{A,B}|^2$ vs. $r/R$) associated to the wavefunctions.
Lower panels:
Plots for $R=10\,\text{nm}$, $R/l_B=3$, corresponding to the region where only $\tau=-1$ levels are present and different values of $j$. Left plot: $j=\{-1/2,-3/2,-5/2,-7/2,-9/2\}$ corresponding to A (B) sublattice curves ordered from bottom (left) to top (right). Right plot: $j=\{-11/2,-13/2,-15/2,-17/2,-19/2\}$ corresponding to A (B) sublattice curves ordered from top (left) to bottom (right).
}
\label{wfwithB}
\end{figure}

\section{Wavefunctions of the bound-states for $B=0$}
\label{appenB=0}

In the case of $B=0$, Eq. (\ref{generalsolution}) reduces to
\begin{align}
&
\chi_\sigma^{\tau}(r)
=
\left\{
\begin{array}{cc}
\delta_{\sigma}J_{j-\sigma/2}\left[k(m_1)r
\right] & \text{for } r\leq R\;,\\
\gamma_{\sigma}H^{(1)}_{j-\sigma/2}\left[i\,k(m_2)r
\right] & \text{for } r>R\;.
\end{array}
\right.
\label{chisnoB}
\end{align}
In turn,
\begin{align}
k(m)\equiv\sqrt{|(\tau E-mv_F^2)(\tau E+mv_F^2)|}/\hbar v_F
\end{align}
and $J_{j-\sigma/2}$ and $H^{(1)}_{j-\sigma/2}$ are the Bessel functions and the Hankel functions of the first kind as defined in Ref. \onlinecite{abramowitz1970handbook}.
The ratio $\delta_\sigma/\gamma_\sigma$ is calculated analogously to $\alpha_\sigma/\beta_\sigma$ in App. \ref{appenBneq0}.

The continuity condition for the wavefunction yields in this case
\begin{equation}
i\,\tau\,\text{sg}(E)\eta(m_2)
J_{j-1/2}\left[k(m_1)R\right]
H^{(1)}_{j+1/2}\left[i\,k(m_2)R\right]
=
\eta(m_1)
J_{j+1/2}\left[k(m_1)R\right]
H^{(1)}_{j-1/2}\left[i\,k(m_2)R\right]\;, \label{Eboundap}
\end{equation}
which again gives the allowed energies of the bound-states. We also defined
\begin{align}
\eta(m)\equiv\sqrt{\left|\frac{\tau E-mv_F^2}{\tau E+mv_F^2}\right|}\;.
\end{align}
The considerations regarding the electron-hole symmetry which are related to the $\tau E$ dependence also apply here, see App. \ref{appenBneq0}.
Moreover, the properties of Bessel and Hankel functions
\begin{align}
J_{-n}(z)=(-1)^nJ_{n}\;,\;\ H_{-\nu}^{(1)}(z)=e^{\nu\pi i}H_{\nu}^{(1)}(z)\;,
\end{align}
together with
\begin{align}
\frac{\eta(m_1)}{\eta(m_2)}\xrightarrow{\tau \to -\tau}\frac{\eta(m_2)}{\eta(m_1)}
\end{align}
allow to prove the double degeneracy of levels, i.e., a solution with $(E,\tau,j)$ implies the existence of another with $(E,-\tau,-j)$.
In contrast with the electron-hole symmetry, the substitutions of Eqs. (\ref{change1}) and (\ref{change2}) in (\ref{Eboundap}) (yielding a PW-QD) do not lift this degeneracy. 
As we mention in the main text and show in Figs. \ref{Splitting} and \ref{SpectrumB}, a splitting happens when a magnetic field is applied. Plots of several wavefunctions and their corresponding radial probabilities are shown in Fig. \ref{wfwithoutB}.

\begin{figure}
\begin{center}
\includegraphics{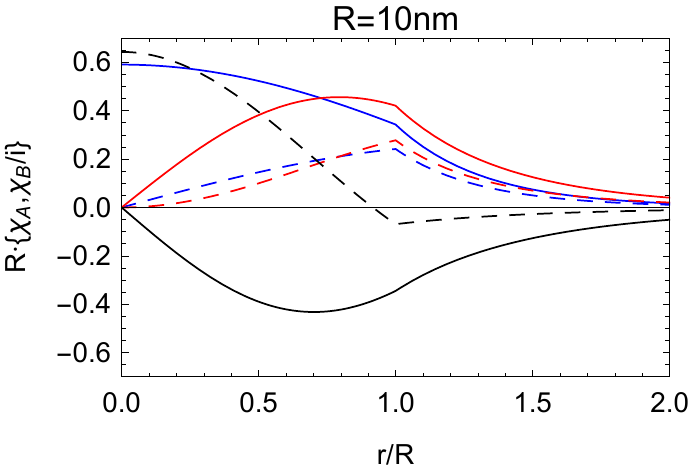}
\hspace{1cm}
\includegraphics{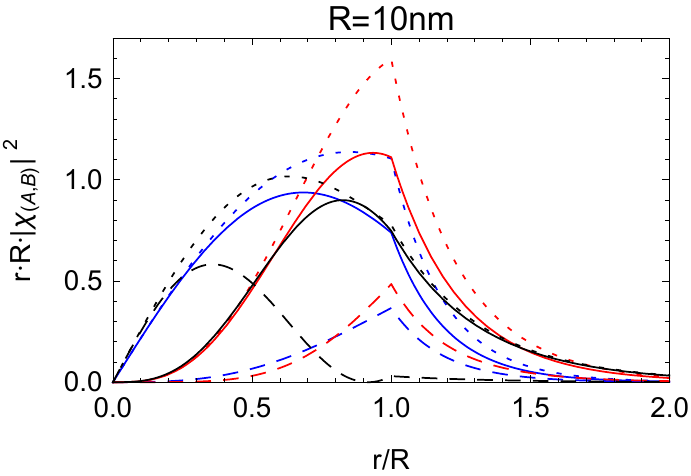}
\end{center}
\caption{(color online). Components of the wavefunction (left hand side), Eqs. (\ref{chisnoB}) and (\ref{wfB}), and their associated radial probabilities (right hand side) for $\tau=1$ and $j=1/2$ (blue), $j=3/2$ (red) and $j=-1/2$ (black). Solid (dashed) lines correspond to the A (B) sublattice, and dotted lines to the total radial density probability. A wavefunction with $\tau=-1$ and $j=-1/2$ is also plotted in Fig. \ref{wfwithB} (right hand side of upper panel).}
\label{wfwithoutB}
\end{figure}

\end{widetext}

\section{Tight binding in a lattice of MP-QDs}
\label{appenB}

\begin{figure}
\begin{center}
\includegraphics{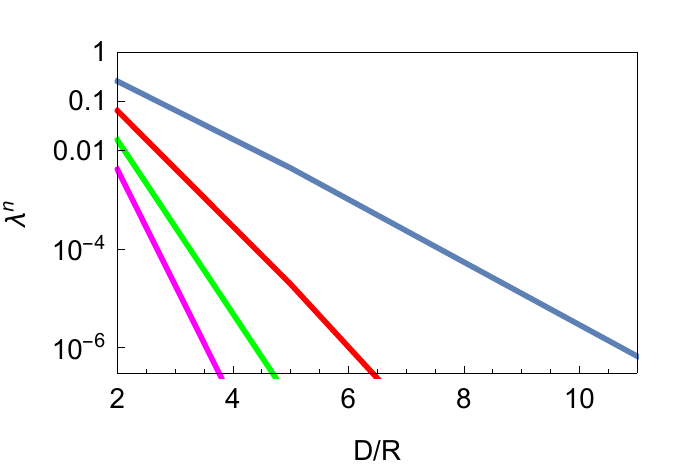}
\end{center}
\caption{(color online). Overlap between wavefunctions belonging to wells whose centers lie at a distance $D$. $n=1,2,3,4$ correspond to blue, red, green and magenta.}
\label{Solape}
\end{figure}

Our aim is to construct a tight binding model for a system in which the overlap $\lambda$ of neighboring wavefunctions cannot be neglected.
This is motivated by Fig. \ref{Solape}, which plots the overlap for a couple of quantum dots as a function of the distance.
As a start point, we will discard terms which are cubic or of higher order in $\lambda$. Actually, we will show that $O(\lambda^2)$ terms will be significant in packed lattices of quantum dots.

Let $\ket{n}$ be the wavefunction of a particular state of a single well located at a certain position. The set $S$ containing all $\ket{n}$ kets respective to every lattice site and every energy level is not orthonormal because the overlap $\bracket{n}{n'}$ between neighbors is not negligible. However, the Gram-Schmidt algorithm allows us to obtain an orthonormal basis $S'$ by linear combinations of the vectors belonging to $S$. Denoting the elements of $S'$ by $\ket{f_n}$, we can write the identity as
\begin{align}
I=\sum_n\ket{f_n}\bra{f_n}\;. \label{idfn}
\end{align}
Carrying out the algorithm to obtain $\ket{f_n}$, the unity-operator in terms of the $S$-basis vectors only including correction terms to lowest orders reads
\begin{align}
I&=
\sum_{n}
\ket{n}\bra{n}
-
\sum_{j\neq n}\bracket{j}{n}\ket{j}\bra{n}\;. \label{id2ndorder}
\end{align}
To express the Hamiltonian in the original basis $S$, we will use this representation of the identity operator.

Let us apply the aforementioned procedure to a set of $N$ MP-QDs whose centers are located at $\vec{R}_i$ with $i=\{1,\dots,N\}$. We will consider wells that only have one valence and conduction bound-state with energy $\pm E$ ($E>0$) and $j=\pm 1/2$ (cf. Fig. \ref{SpectraB0}). Therefore, we can univocally label the $S$ states by $\ket{{\pm\vec{R}_i}}$.


We define the following parameters, describing the hopping processes depicted in Fig. \ref{pozoshoppings}:
\begin{align}
\lambda_{ij}^\pm&=\bra{\pm\vec{R}_i}{\pm\vec{R}_j}\rangle\;,
\label{parini}
\\
\xi^{\pm}&=\bra{\pm\vec{R}_i}\Delta U_i\ket{{\pm\vec{R}_i}}\;,
\\
\eta^\pm&=\bra{\pm\vec{R}_i}\Delta U_i|{\mp\vec{R}_i}\rangle\;,
\\
\kappa_{ij}^\pm&=\bra{\pm\vec{R}_i}\Delta U_j|{\pm\vec{R}_j}\rangle\;,
\\
\mu_{ij}^\pm&=\bra{\pm\vec{R}_i}\Delta U_j|{\mp\vec{R}_j}\rangle\;. \label{parfin}
\end{align}

Assuming inversion symmetry for the array under consideration and making use of the properties of the wavefunctions, the following identities can be proved:
\begin{align}
&
\{\lambda^{\pm}_{ij},\xi^{\pm},\kappa^{\pm}_{ij}\}\in\mathbb{R}\;,\ 
\lambda^{+}_{ij}=\lambda^{-}_{ij}\equiv\lambda_{ij}\;,\ \lambda_{ij}=\lambda_{ji}\;,\\
&
\xi^+=-\xi^-\;,\ 
\eta^{\pm}=0\;,\ 
\kappa^{+}_{ij}=-\kappa^{-}_{ij}\;,\ 
\kappa^{\pm}_{ij}=\kappa^{\pm}_{ji}\;,\\
&
\mu^{\pm}_{ij}\equiv\mu^{\pm}(\vec{R}_i-\vec{R}_j)=\mu^{\pm}(|\vec{R}_j-\vec{R}_i|\hat{x})e^{\pm i\theta_{\vec{R}_j-\vec{R}_i}}\;, \label{property3} \\
&
\mu^{\pm}(|\vec{R}_j-\vec{R}_i|\hat{x})\in\mathbb{R}\;,\ \mu^{+}_{ij}=-{\mu^{-}_{ij}}^*.
\end{align}
$\hat{x}$ is the unitary vector in the $x$ direction, $\theta_{\vec{R}_j-\vec{R}_i}$ is the angle between $\vec{R}_j-\vec{R}_i$ and $\hat{x}$ and $h(\vec{R}_i-\vec{R}_j)\equiv h_{ij}$, where $h=\{\lambda^{\pm},\kappa^{\pm},\mu^{\pm}\}$.
In turn,
\begin{align}
\Delta U_j=H-H_{\vec{R}_j},
\end{align}
$H=\sum_iH_{\vec{R}_i}$ being the total Hamiltonian and $H_{\vec{R}_j}$ the Hamiltonian of a single MP-QD centered at $\vec{R}_j$.
$\Delta U_j$ accounts then for the influence of the lattice on the Hamiltonian of an isolated dot and results in the hopping of electrons between different wells.

The dependence of Eqs. (\ref{parini})-(\ref{parfin}) on $D/R$, where $D$ is the distance between the centers of the dots, is plotted in Fig. \ref{Eorders} for a double MP-QD.
It can be seen that $\{\mu,\kappa\}<\lambda(m_2-m_1)v_F^2$ and $\xi<\lambda^2(m_2-m_1)v_F^2$, which will be taken into account when discarding terms of greater order than $\lambda^2$ in subsequent calculations.

Under these considerations and with Eq. (\ref{id2ndorder}), the Hamiltonian acting on $|{\pm\vec{R}_m}\rangle$ can be expressed as
\begin{align}
&
H\ket{{\pm\vec{R}_m}}=
\left(H_{\vec{R}_m}+I\cdot\Delta U_m\right)\ket{{\pm\vec{R}_m}}=
\notag
\\
&
=
\left[
\pm|E|+\xi^{\pm}
-
\sum_{i\neq m}
\lambda_{mi}
\kappa_{im}^{\pm}
\right]
\ket{{\pm\vec{R}_m}}\notag
\\
&
+\sum_{j\neq m}
\left[
\kappa_{jm}^{\pm}
-
\xi^{\pm}\lambda_{jm}
-
\sum_{i\neq \{j,m\}}
\lambda_{ji}
\kappa_{im}^{\pm}
\right]
\ket{{\pm\vec{R}_j}} \notag
\\
&
+
\sum_{j\neq m}
\left[
\mu_{jm}^{\pm}
-
\sum_{i\neq\{j,m\}}
\lambda_{ji}
\mu_{im}^{\pm}
\right]
\ket{{\mp\vec{R}_j}}
\notag
\\
&
-
\sum_{j\neq m}
\lambda_{mj}
\mu_{jm}^{\pm}
\ket{{\mp\vec{R}_m}} \label{Hvec}
.
\end{align}

\begin{figure*}
\begin{center}
\begin{minipage}{\textwidth}
\raisebox{-0.5\height}{\includegraphics{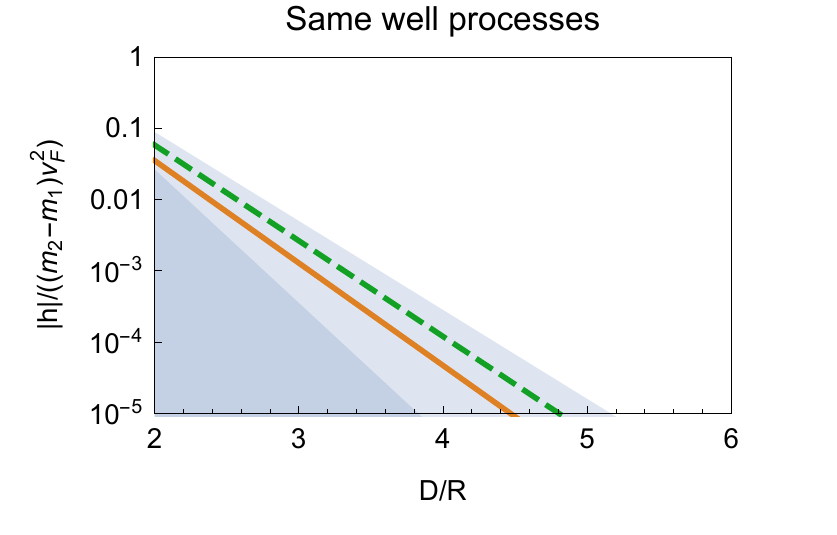}}
\raisebox{-0.3\height}{\includegraphics{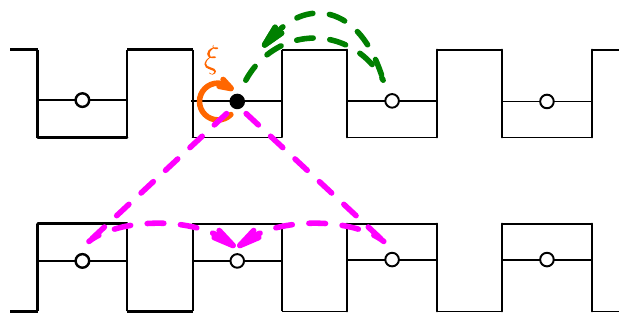}}
\end{minipage}
\begin{minipage}{\textwidth}
\raisebox{-0.5\height}{\includegraphics{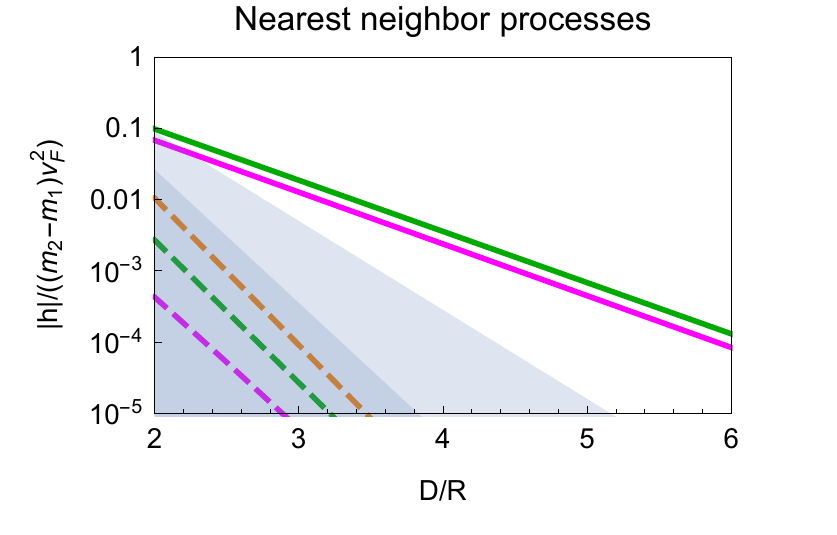}}
\raisebox{-0.3\height}{\includegraphics{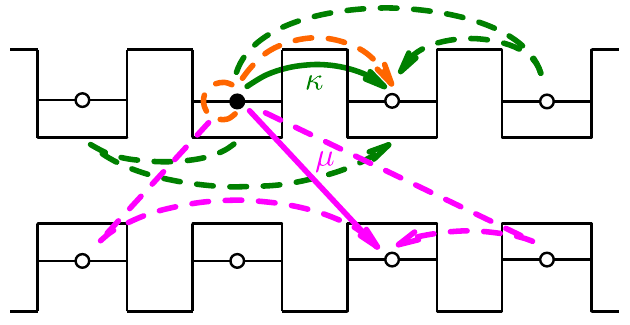}}
\end{minipage}
\begin{minipage}{\textwidth}
\raisebox{-0.5\height}{\includegraphics{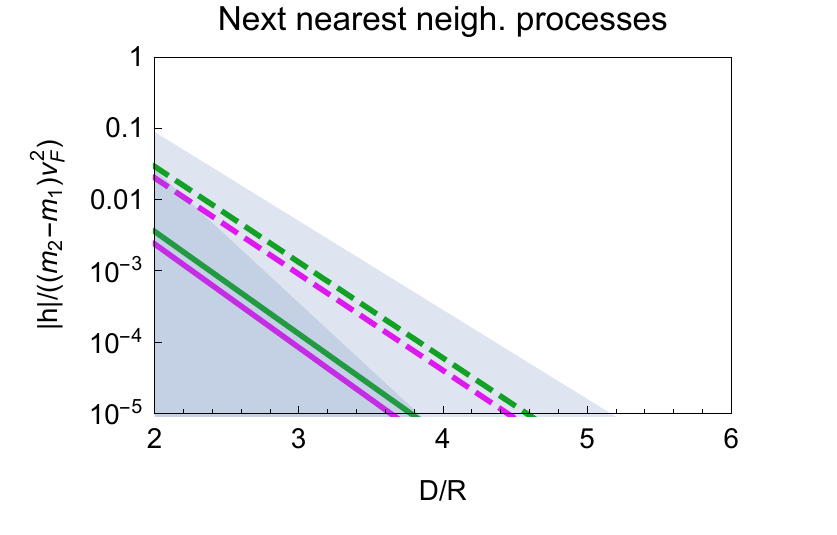}}
\raisebox{-0.3\height}{\includegraphics{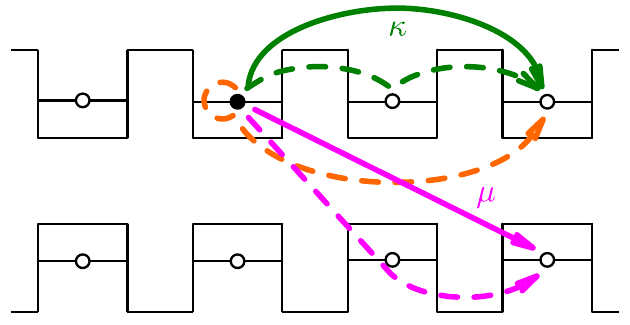}}
\end{minipage}
\end{center}
\caption{(color online). Analogous to Fig. \ref{Eorders}.
Hopping processes to the same well, to a nearest neighbor or to a next-nearest neighbor are analyzed separately.
The sketches on the right give the legend of the plots on the left.
Solid lines correspond to direct processes, and dashed lines to processes assited by $\lambda$ together with a direct hopping, i.e., terms of the kind $\sum_{i,j,m}
\lambda_{ji}
h_{im}^{\pm}$ appearing in Eq. (\ref{Hvec}) ($h=\{\xi,\kappa,\mu\}$).
When in each right outline several processes are labeled by a single linestyle, the associated curve on the left plot corresponds to the sum of their probability amplitues.
In the upper left plot, the magenta processes do not appear because the sum of their amplitudes vanishes due to Eq. (\ref{property3}).
In the bottom left plot, the amplitude of the orange process is not identically zero, but so small that lies outside the plot range.
For shadowed regions, see Fig. \ref{Eorders}.
}
\label{neighbors}
\end{figure*}

Eq. (\ref{Hvec}) gives the matrix elements of $H$ in the $S$ basis for a still unspecified geometry of the quantum dot set. 
This general result can be applied to different systems.
The simplest consists in only two coupled MP-QDs.
Its spectrum, with the individual energy levels splitted, appears in Fig. \ref{Eorders}.
One can see there that for $D\lesssim 4R$, the influence of $O(\lambda)$-terms is significant, although there is not a great difference between neglecting $\lambda^2$- and $\lambda^3$-terms even at small distances between the wells.
Second order processes, however, will be more relevant in lattices due to the assistance of next-nearest neighbor hopping processes, see the following discussion and Fig. \ref{neighbors}.

For a periodic system of  MP-QDs, it is more convenient to work in a Fourier transformed basis defined by
\begin{align}
\ket{{\pm\vec{R}}}=
\frac{1}{2\pi}
\int d^2k\ 
e^{i\vec{k}\cdot\vec{R}}\ket{{\pm\vec{k}}}.
\label{Fourierbasis}
\end{align}
Inserting Eq. (\ref{Fourierbasis}) in Eq. (\ref{Hvec}), the Hamiltonian can be expressed in block diagonal form. The block respective to $\vec{k}$ reads
\begin{align}
&
H(\vec{k})\equiv
\left[
\begin{array}{cc}
\bra{{-\vec{k}}}H\ket{{-\vec{k}}}		&
\bra{{-\vec{k}}}H\ket{{+\vec{k}}}		\\
\bra{{+\vec{k}}}H\ket{{-\vec{k}}}		& 
\bra{{+\vec{k}}}H\ket{{+\vec{k}}}
\end{array}
\right]
=\notag
\\
&
=
\left[
\begin{array}{cc}
h_{1}^{-}+\sum_{\vec{\delta}} h_{2,\vec{\delta}}^{-}e^{i\vec{k}\cdot\vec{\delta}}		& \sum_{\vec{\delta}} h_{3,\vec{\delta}}^{-}e^{i\vec{k}\cdot\vec{\delta}}		\\
\sum_{\vec{\delta}} h_{3,\vec{\delta}}^{+}e^{i\vec{k}\cdot\vec{\delta}}	& h_{1}^{+}+\sum_{\vec{\delta}} h_{2,\vec{\delta}}^{+}e^{i\vec{k}\cdot\vec{\delta}}		
\end{array}
\right]. \label{Hkfinal}
\end{align}
Dropping the subindicies, $\vec{\delta}$ is defined by
\begin{align}
\vec{\delta}=\vec{R}_j-\vec{R}_m
\end{align}
choosing the adequate $j$ and $m$. We have also defined correspondingly
\begin{align}
&
h_{1}^{\pm}=
\pm|E|
+
\xi^{\pm}
-
\sum_{i\neq m}
\lambda_{mi}
\kappa_{im}^{\pm} \label{h1}
\;,
\\
&
h_{2,\vec{\delta}}^{\pm}=\kappa_{jm}^{\pm}
-
\xi^{\pm}\lambda_{jm}
-
\sum_{i\neq \{j,m\}}
\lambda_{ji}
\kappa_{im}^{\pm}\;,\label{h2}
\\
& h_{3,\vec{\delta}}^{\pm}=\mu_{{jm}}^{\pm}
-
\sum_{i\neq m}
\lambda_{ji}
\mu_{im}^{\pm}. \label{h3}
\end{align}
The following identities, the last of which guarantees the hermiticity of the Hamiltonian, can be proved attending to the symmetry of the wavefunctions and $\Delta U_j$:
\begin{align}
h_1^{+}=-h_1^{-};\ 
h_{2,\vec{\delta}}^{+}=-h_{2,\vec{\delta}}^{-};\ 
h_{3,\vec{\delta}}^{+}={h_{3,-\vec{\delta}}^{-}}^*.
\end{align}

We can apply the general result Eq. (\ref{Hkfinal}) to the simplest lattice, namely a one-dimensional chain with a single atom per node.
To do so, we must determine the relevant processes which contribute significantly to Eqs. (\ref{h1})-(\ref{h3}).
That analysis was carried out and is summarized in Fig. \ref{neighbors}.

There, we see that departing from the criterium of neglecting $\lambda^2$, only the direct processes $\kappa$ and $\mu$ between neighboring wells are relevant.
However, as long as the chain lattice parameter becomes close to the diameter of the wells, $O(\lambda^2)$-processes become increasingly more relevant.
In particular, a next-nearest neighbor (nnn) hopping process is assisted by the wavefunction overlap, whereas direct processes to nnn are negligible.

Once we have discarded all irrelevant terms in Eqs. (\ref{h1})-(\ref{h3}), we can diagonalize Eq. (\ref{Hkfinal}) to obtain the bands and the tight binding description is complete. Spectra for various lattice parameters are plotted in Fig. \ref{Figbandas}.

\bibliography{massprofile}

\end{document}